# An ambient denoising method based on multi-channel non-negative matrix factorization for wheezing detection

Antonio J. Muñoz-Montoro[1*], Pablo Revuelta-Sanz[1†], Damian Martínez-Muñoz[2†], Juan Torre-Cruz[2†] and José Ranilla[1†]

[1*]Department of Computer Science, University of Oviedo, Campus de Gijón, Gijón, 33204, Asturias, Spain.
[2]Departament of Telecommunication Engineering, University of Jaen, Campus Cientifico-Tecnologico de Linares, Linares, 23700, Jaen, Spain.

*Corresponding author(s). E-mail(s): munozantonio@uniovi.es;
Contributing authors: revueltapablo@uniovi.es; damian@ujaen.es; jtorre@ujaen.es; ranilla@uniovi.es;
[†]These authors contributed equally to this work.

**Abstract**

In this paper, a parallel computing method is proposed to perform the background denoising and wheezing detection from a multi-channel recording captured during the auscultation process. The proposed system is based on a non-negative matrix factorization (NMF) approach and a detection strategy. Moreover, the initialization of the proposed model is based on singular value decomposition (SVD) to avoid dependence on the initial values of the NMF parameters. Additionally, novel update rules to simultaneously address the multichannel denoising while preserving an orthogonal constraint to maximize source separation have been designed. The proposed system has been evaluated for the task of wheezing detection showing a significant improvement over state-of-the-art algorithms when noisy sound sources are present. Moreover, parallel and high-performance techniques have been used to speedup the execution of the proposed system, showing that it is possible to achieve fast execution times, which enables its implementation in real-world scenarios.

**Keywords:** Non-negative matrix factorization (NMF), Singular value decomposition (SVD), Parallel computing, Wheezing detection, Denoising, Multi-channel

# 1 Introduction

Breath sounds analysis is a relevant technique to assess the state of lungs and other organs that compose the respiratory system. Abnormal sounds overlapping normal respiratory sounds can alert of a respiratory disorder. Thus, one of the main challenges for current biomedicine is the development of new methodologies to process respiratory sounds and to obtain reliable and individualized biomedical information, by using non-invasive techniques. Fortunately, respiratory sounds can be acquired by an easy and non-invasive auscultation procedure, which allows the extraction of relevant information from the lungs and helps to reduce diagnostic time [1, 2]. Therefore, auscultation remains one of the most widely used techniques to detect any lung disease. However, this technique suffers from two main limitations, the high dependence on ambient noise surrounding auscultation, preventing a reliable diagnosis [3], and the high subjectivity of physicians, who sometimes are not able to recognize sounds related to a physiological disorder [4].

Removing ambient noises from auscultation recordings to maximize the reliability of diagnoses has been a hot topic in the biomedical signal processing field during the last decade. In this sense, most of the approaches developed are based on adaptive filtering [5–9] and spectral subtraction [10–13]. Fleeter and Wodicka [7] proposed to apply two adaptive filtering algorithms, least mean square (LMS) and normalized LMS (NLMS), to reduce auscultation noise in an aircraft with reduced adaptation time. Chang and Lai [11] developed a spectral subtraction method based on mel-frequency cepstral coefficients (MFCC), autoregressive (AR) and dynamic time warping (DTW). The algorithm was applied to the lung sound signals under noisy conditions, before the extraction of lung sound features. Emmanouilidou et al. [12] proposed a two-microphone multiband spectral scheme to remove the background noise while preserving the lung sounds to maximize the informative diagnostic value obtained from auscultation. The method analyzes each frequency band in a non-uniform manner and uses prior knowledge of the target sounds to apply a penalty in the spectral domain. On the other hand, recent researches based on non-negative matrix factorization (NMF) approaches have addressed this type of biomedical problem [14–16]. Torre-Cruz et al. [14] proposed an incremental algorithm based on non-negative matrix partial co-factorization (NMPCF) that improves the quality of biomedical sounds captured in auscultated recordings by applying the conventional NMPCF from a multi-channel scenario rather than a single-channel.

On the other hand, one of the diagnoses that are still not correct nowadays due to the subjective evaluation of the physician is the wheezing detection. Wheeze sounds have been considered a reliable indicator of the degree of the bronchial obstruction related to several pulmonary diseases [17], such as asthma, acute bronchitis, bronchiolitis, etc. These sounds are characterized for some specific spectral [18–20] and temporal [18, 20, 21] features. However, wheeze and normal breath sounds are mixed together in the time-frequency domain since both are simultaneously generated by the same airflow [22, 23]. Several strategies have been proposed to deal with wheezing detection in the literature, such as spectral peaks detection [24, 25], Hidden Markov Models [26], Gaussian mixture models [27], auditory modelling [28], wavelet transforms [29], autoregresive models [30], neural networks [31–35], etc. Recently, NMF approches have been also proposed for the wheezing detection obtaining promising results [36, 37].

In this work, a parallel computing system is proposed to address jointly the background denoising and the wheezing detection from a multi-channel recording captured during the auscultation process. Specifically, a two-channel signal model, suitable to distinguish between wheezing and normal breath sounds, is proposed. To deal with this problem, the periodicity principle of the wheezing sounds is exploited. The proposal is an efficient and fast implementation that is able to perform the decomposition of the input mixture by using a NMF approach and to perform the wheezing detection by using a sparse descriptor. Unlike previous NMF-based methods [14, 37] where incremental and recursive algorithms with high computational cost were proposed, the main contribution of this work is a signal decomposition method that allows to perform jointly the denoising and the wheezing detection of an input respiratory signal in a single step, reducing thus the computational cost of the algorithm. Moreover, we have introduced an initialization method based on singular value decomposition (SVD) to avoid dependence on the initial values of the NMF parameters. The tested scenarios show that, combining parallel and high-performance techniques, our proposal can be applied to real scenarios.

According to the best of our knowledge, there has not yet been presented a holistic and flexible system that addresses jointly both problems on parallel shared-memory systems. As a proof of concept, some experiments are performed on a dataset of real-world audio samples, showing promising results in terms of computational and reliability.

The structure of the rest of the article is as follows. In Section 2, we present the problem formulation and briefly review the multi-channel denoising and wheezing detection based on NMF approaches. The proposed system is presented in Section 3. Experimental results are shown in Section 4. Finally, we summarize the work and discuss the future perspectives in Section 5.

## 2 Background

### 2.1 Problem formulation

The problem considered in this work is the detection of wheezing sounds from the two-channel signal recorded by a digital stethoscope. Thus, the observed signals can be formulated as

$$x_I(n) = s(n) + v(n) \tag{1}$$
$$x_E(n) = v(n) \tag{2}$$

where the time-domain sample index is denoted by $n$, $x_E(n)$ is the external signal captured by the external microphone that records the ambient noise signal $v(n)$, and $x_I(n)$ represents the internal signal captured by the stethoscope and composed by the biomedical sounds $s(n)$ from the subject and the ambient noises $v(n)$ surrounding the subject that are still heard inside the human body.

From (1), the short-time Fourier transform (STFT) of both external and internal signals can be expressed as

$$X_I(f, t) = S(f, t) + V(f, t) \tag{3}$$
$$X_E(f, t) = V(f, t) \tag{4}$$

where $X_I(f, t)$, $X_E(f, t)$, $S(f, t)$ and $V(f, t)$ represent the magnitude spectrograms of $x_I(n)$, $x_E(n)$, $s(n)$ and $v(n)$, respectively. Note that $f \in [1, F]$ and $t \in [1, T]$ denote the frequency bin and time frame indices, respectively. Collecting $F$ frequency bins and $T$ time frames, we define the magnitude spectrogram matrices $\mathbf{X}_I \in \mathbb{R}_+^{F \times T}$, $\mathbf{X}_E \in \mathbb{R}_+^{F \times T}$, $\mathbf{S} \in \mathbb{R}_+^{F \times T}$ and $\mathbf{V} \in \mathbb{R}_+^{F \times T}$, where $\mathbf{X}_I = [\mathbf{x}_I(1), \ldots, \mathbf{x}_I(t), \ldots, \mathbf{x}_I(T)]$ and $\mathbf{x}_I(t) = [X_I(1, t), \ldots, X_I(f, t), \ldots, XI(F, t)]^T$. $\mathbf{x}_E(t)$, $\mathbf{s}(t)$ and $\mathbf{v}(t)$ are defined in the same manner as $\mathbf{x}_I(t)$. Finally, $\mathbf{X}_E$, $\mathbf{S}$ and $\mathbf{V}$ are defined similarly to $\mathbf{X}_I$.

### 2.2 Multi-channel denoising based on NMPCF

During the auscultation, noisy environments can prevent a proper diagnosis of subjects. Ambient noise often overlap with biomedical sounds, thereby polluting valuable clinical information.

This problem is tackled in [14]. The authors proposed an algorithm for denoising the two-channel biomedical signal captured by a stethoscope in very noisy environments. The proposal consists of an incremental algorithm based on a Non-negative Matrix Partial Co-Factorization (NMPCF) approach. The decomposition model presented was given by

$$\mathbf{X}_I \approx \hat{\mathbf{X}}_I = \hat{\mathbf{S}} + \hat{\mathbf{V}} = \underbrace{[\mathbf{B}_S \ \mathbf{B}_V]}_{\mathbf{B}} \underbrace{\begin{bmatrix} \mathbf{G}_S \\ \mathbf{G}_V \end{bmatrix}}_{\mathbf{G}} = \mathbf{B}_S \mathbf{G}_S + \mathbf{B}_V \mathbf{G}_V \quad (5)$$

$$\mathbf{X}_E \approx \hat{\mathbf{X}}_E = \hat{\mathbf{V}} = \mathbf{B}_V \mathbf{H}_V \quad (6)$$

where $\hat{\mathbf{X}}_I \in \mathbb{R}_+^{F \times T}$ and $\hat{\mathbf{X}}_E \in \mathbb{R}_+^{F \times T}$ are the estimated magnitude spectrogram of the internal and external signals, respectively. $\mathbf{B}_S \in \mathbb{R}_+^{F \times K_S}$ and $\mathbf{G}_S \in \mathbb{R}_+^{K_S \times T}$ are the bases and gains matrices of the biomedical sounds and $\mathbf{B}_V \in \mathbb{R}_+^{F \times K_V}$ and $\mathbf{G}_V \in \mathbb{R}_+^{K_V \times T}$ are the bases and gains matrices of the ambient noise, all of them obtained from the internal signal. The parameters $K_S$ and $K_V$ denote the number of bases related to the biomedical sounds and the ambient noise, respectively. $\mathbf{H}_V \in \mathbb{R}_+^{K_V \times T}$ is the gain matrix of the ambient noise obtained from the external signal.

Note that this model allows to factorize jointly $\mathbf{X}_I$ and $\mathbf{X}_E$. In this way, the spectral patterns of the ambient noise are shared in the same dictionary $\mathbf{B}_V$ assuming that the noise are simultaneously active in both spectrograms. On the other hand, $\mathbf{B}_S$ represents the spectral patterns of the biomedical sounds that are only active in the internal magnitude spectrogram $\mathbf{X}_I$.

The authors proposed to deal with the factorization by minimizing the generalized Kullback-Leibler $D_{KL}(\mathbf{Z}|\hat{\mathbf{Z}})$ divergence [14]. The update rules for this signal model can be found in [14].

To improve the biomedical signal estimation, the authors proposed an incremental algorithm that runs NMPCF several times. The algorithm relies on running NMPCF taking as input the biomedical signal estimate from the previous incremental iteration. In this way, all spectral patterns associated to ambient noises are extracted from the target signal. In this approach, the external signal is fixed for all incremental iterations since it is only composed by ambient noises.

Finally, the reconstruction of the target signals is carried out by using the inverse overlap-add STFT combined with a soft-filter strategy. Thus, the estimated parameters are used to predict the spectrograms of the biomedical and ambient noises signals by

$$\hat{s}(t) = iSTFT \left( \mathbf{X}_I \odot \frac{|\mathbf{B}_S \mathbf{G}_S|^2}{|\mathbf{B}_S \mathbf{G}_S|^2 + |\mathbf{B}_V \mathbf{G}_V|^2} \right) \quad (7)$$

$$\hat{v}(t) = iSTFT \left( \mathbf{X}_I \odot \frac{|\mathbf{B}_V \mathbf{G}_V|^2}{|\mathbf{B}_S \mathbf{G}_S|^2 + |\mathbf{B}_V \mathbf{G}_V|^2} \right) \quad (8)$$

where $\odot$ represents the element-wise product. The reader can refer to [14] for more details.

### 2.3 Wheezing detection based on ONMF

The detection of wheeze sounds from single-channel audio mixtures based on orthogonal NMF (ONMF) (i.e., the orthogonality constraint integrated into the NMF factorization procedure) has been addressed in [37]. The authors propose a recursive algorithm based on the periodicity principle to discriminate between wheezing and normal breath spectral templates assuming that a wheeze sound exhibits a strongly periodic or tonal nature.

The decomposition model presented in that work is given by

$$\mathbf{X} \approx \hat{\mathbf{X}} = \mathbf{BG} \qquad (9)$$

where $\hat{\mathbf{X}} \in \mathbb{R}_+^{F \times T}$ is the estimated magnitude spectrogram of the input signal, $\mathbf{B} \in \mathbb{R}_+^{F \times K}$ is the bases matrix and $\mathbf{G} \in \mathbb{R}_+^{K \times T}$ is the activations matrix. $K$ is the number of bases or components.

The factorization in (9) is carried out by minimizing a global objective function $D(\mathbf{X}|\hat{\mathbf{X}})$ which integers the Kullback-Leibler divergence $D_{KL}(\mathbf{X}|\hat{\mathbf{X}})$ and the orthogonality constraint $\phi(\mathbf{B})$ [37]. Thus, the update rules were defined as follows

$$\mathbf{B} \leftarrow \mathbf{B} \odot \sqrt{\frac{\mathbf{XG}^T}{\mathbf{BB}^T\mathbf{XG}^T}} \qquad (10)$$

$$\mathbf{G} \leftarrow \mathbf{G} \odot \frac{\mathbf{B}^T\mathbf{X}}{\mathbf{BB}^T\mathbf{G}} \qquad (11)$$

where $\mathbf{B}$ and $\mathbf{G}$ are initialized as random positive matrices.

Then, Torre-Cruz et al. [37] proposed a clustering process to classify the bases $\mathbf{B}$ computed in the ONMF decomposition. This clustering is based on the periodic and non-periodic nature of a wheeze and normal breath sound. The sparse descriptor Gini index in the frequency domain is used to distinguish between bases with a high periodicity and bases with a low periodicity.

This factorization and clustering procedure is repeated recursively to improve the estimated wheezing spectrogram. At each recursive iteration $i$, the wheezing spectrogram is the input of recursive iteration $i+1$ in order to obtain a new set of ONMF bases that must be reclustered into wheezing bases or normal breath bases according to the discrimination performed by the sparse descriptor Gini index. Then, the authors define a threshold to determine the optimal iteration $i_o$ that implies the stop of the recursive process.

## 3 Proposed algorithm for multi-channel wheezing detection

In this work, a multi-channel system based on NMF for wheezing detection is presented. Specifically, we propose a two-channel signal model suitable to distinguish between wheezing and normal breath sounds. To deal with this problem, the periodicity principle of the wheezing sounds is exploited. For this purpose, an efficient and fast implementation has been developed that is able to perform the decomposition of the input mixture using a NMF approach and to perform the wheezing detection using a sparse descriptor. As a result, we propose a software solution that satisfies two essential requirements: mobility and real-time scenarios. Thus, our design takes into account the low memory resources and low computational power of cheap and handheld devices, what can allow an easy implementation in the medical services.

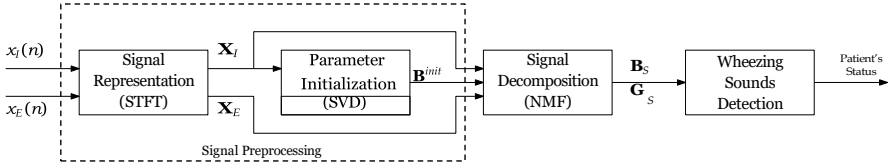

**Fig. 1**: Block diagram of the proposed algorithm.

The block diagram of the proposed algorithm is depicted in Fig. 1. As can be observed, the issue has been decomposed into three main stages: signal preprocessing, signal decomposition and wheezing detection. The following subsections detail the main function and the procedure of each stage.

## 3.1 Signal preprocessing

As can be seen in Fig. 1, the preprocessing stage must be computed beforehand and consists of two successive steps: signal representation and parameter initialization.

Considering the mixture model described in Sect. 2.1, the first step consists of computing the STFT of the audio mixtures $x_I(n)$ and $x_E(n)$ (see details in Sect. 2.1). Then, we propose to use an effective initialization step for the NMF approach with the aim of reducing its computational complexity and improving the factorization. NMF is a powerful unsupervised learning method that extracts meaningful nonnegative features from an observed nonnegative data matrix. However, the result obtained by this algorithm always depends on the initial values of the NMF parameters, due to the existence of local minima. To solve this problem, we propose a unique initialization for the NMF bases parameters based on SVD [38]. Thus, any random values or hyperparameters are not required. The main feature of this approach is that the NMF algorithm converges to the same solution while rapidly providing an approximation with error almost as good as that obtained via the deployment of alternative initialization schemes [38].

In this work, we propose to initialize the bases $\mathbf{B}_S$ and $\mathbf{B}_V$ in (5) and (6) by using the left singular matrix obtained by SVD. The basic property of the SVD relies on the fact that every matrix $\mathbf{X} \in \mathbb{R}_+^{m \times n}$ of rank $R$ can be expressed as the sum of $R$ leading singular factors,

$$\mathbf{X} = \sum_{r=1}^{\bar{R}} \alpha_r \mathbf{u}_r \mathbf{v}_r^T, \qquad (12)$$

where $\alpha_1 \geq \cdots \geq \alpha_R \geq 0$ are the nonzero singular values of $\mathbf{X}$ and $\{\mathbf{u}_r, \mathbf{v}_r\}_{r=1}^{R}$ the corresponding left and right singular vectors. With the singular values in a diagonal matrix $\xi$ and the corresponding singular vectors forming the columns of two orthogonal matrices $\mathbf{U}$ and $\mathbf{V}$, the SVD decomposition can be expressed as

$$\mathbf{X} = \mathbf{U}\xi\mathbf{V}^T \tag{13}$$

Note that the rank of this decomposition is $R$. In this regards, we propose to apply the SVD algorithm to $\mathbf{X}_I$ and to initialize bases and activations implementing a low rank approximation following the approach described in [39] to reduce the size of the NMF parameters and, therefore, the computational complexity of the signal decomposition stage. Thus, the rank-$k$ approximation of $\mathbf{X}_I$ can be formulated as

$$\mathbf{X}_I^{(k)} = \mathbf{U}\begin{pmatrix}\xi^{(k)}\\0\end{pmatrix}\mathbf{V}^T \tag{14}$$

where $\xi^{(k)}$ is a diagonal matrix compounded by the first $k$ singular values. Then, $\left|\mathbf{U}\begin{pmatrix}\xi^{(k)}\\0\end{pmatrix}^{1/2}\right|$ is used to initialize the bases, while the activation matrices are initialized with random values for the factorization stage. As can be observed, in this case the number of bases for the NMF algorithm is $k$, where $k \ll R$.

The computational complexity of the preprocessing stage is mainly determined by the STFT and SVD computation. For the STFT implementation, the FFTW package [40] has been used. The overall complexity of the sequential version to compute the magnitude spectrogram of the input is given as

$$O(T(F\ \log_2(F))), \tag{15}$$

where $F$ is the total number of frequency bins and $T$ is the number of frames. For the parallel design of the STFT computation, we have exploited parallel and worksharing constructors of OpenMP [41] and we have chosen a coarse-grained parallelism for our implementation (i.e., running several sequential FFT simultaneously by using the proper OpenMP directives (*pragma omp parallel*)). This approach provided the best performance. Therefore, the parallel complexity can be approximated by

$$O\left(\frac{T}{p}(F\ \log_2(F))\right), \tag{16}$$

where $p$ is the total number of used cores.

Regarding the parameter initialization step, the SVD algorithm has been implemented using the LAPACK implementation based on blocked Householder transformations presented in [42]. Thus, the SVD of a matrix $\mathbf{A}$ is firstly obtained by a bidiagonalization method that consists of applying orthogonal matrices on both the left and right sides of $\mathbf{A}$. These two orthogonal matrices are represented as products of elementary Householder reflectors. After the bidiagonal reduction, LAPACK solves the bidiagonal SVD using QR iteration by applying the Givens rotations. Finally, the singular vectors of **A** are obtained. The overall complexity of the described implementation for the SVD algorithm is $O(T^3)$ according to [42].

The pseudocode of this stage is detailed in lines (1)-(4) of Algorithm 1.

## 3.2 Signal decomposition

As can be observed in Fig. 1, the second main stage of the proposal is the signal decomposition. This stage iteratively decomposes the multichannel mix signal using an NMF-based approach.

In this work, we propose to extend the decomposition signal model presented in Sect. 2.2 to deal with the wheezing detection of the multi-channel input signal. As the model in (5)-(6) provides an approximation factorization of the input spectrograms, the aim is to find a factorization that optimizes a given goodness-of-fit measure called cost function. Thus, the corresponding NMF problem can be rewritten as a constrained optimization problem for a given cost function C. In this way, we propose to apply the gradient descend algorithm to minimize the following cost function between the observed and the estimated signal spectrograms

$$\mathsf{C} = D_{KL}(\mathbf{X}\|\hat{\mathbf{X}}) + D_{KL}(\mathbf{Y}\|\hat{\mathbf{Y}}) + \gamma_o(\mathbf{B}_S) \tag{17}$$

where $D_{KL}(\mathbf{X}\|\hat{\mathbf{X}})$ is the generalized Kullback–Leibler divergence [43] and $\gamma_o(\mathbf{B}_S)$ is the orthogonality constraint that ensure that the estimated bases $\mathbf{B}_S$ are as dissimilar (orthogonal) as possible. Both functions are defined as follows

$$D_{KL}(\mathbf{X}|\hat{\mathbf{X}}) = \sum_{ft} \mathbf{X}\log\frac{\mathbf{X}}{\hat{\mathbf{X}}} - \mathbf{X} + \hat{\mathbf{X}} \tag{18}$$

$$\gamma_o(\mathbf{B_S}) = \beta_{\gamma_o}\sum_{K_S K_S} \mathbf{B}_S\mathbf{B}_S^\mathrm{T} - \mathrm{Trace}(\mathbf{B}_S\mathbf{B}_S^\mathrm{T}) \tag{19}$$

where the Trace(**Z**) operator computes the sum of diagonal elements of a square matrix **Z**, $^\mathrm{T}$ is the transpose operator and $\beta_{\gamma_o}$ is a constant factor used to calibrate the importance of the orthogonality constraint in the decomposition process.

In this way, the update rules to estimate the signal model parameters are given by

$$\mathbf{B}_S \leftarrow \mathbf{B}_S \odot \frac{(\mathbf{X}\oslash\hat{\mathbf{X}})\mathbf{G}_S^\mathrm{T} + \beta_{\gamma_o}\mathbf{B}_S}{\mathbf{1}_{FT}\mathbf{G}_S^\mathrm{T} + \beta_{\gamma_o}\mathbf{B}_S\mathbf{1}_{K_S K_S}} \tag{20}$$

$$\mathbf{B}_V \leftarrow \mathbf{B}_V \odot \frac{(\mathbf{X}\oslash\hat{\mathbf{X}})\mathbf{G}_V^\mathrm{T} + (\mathbf{Y}\oslash\hat{\mathbf{Y}})\mathbf{H}_V^\mathrm{T}}{\mathbf{1}_{FT}\mathbf{G}_V^\mathrm{T} + \mathbf{1}_{FT}\mathbf{H}_V^\mathrm{T}} \tag{21}$$

$$\mathbf{G}_S \leftarrow \mathbf{G}_S \odot \frac{\mathbf{B}_S^T(\mathbf{X} \oslash \hat{\mathbf{X}})}{\mathbf{B}_S^T \mathbf{1}_{FT}} \tag{22}$$

$$\mathbf{G}_V \leftarrow \mathbf{G}_V \odot \frac{\mathbf{B}_V^T(\mathbf{X} \oslash \hat{\mathbf{X}})}{\mathbf{B}_V^T \mathbf{1}_{FT}} \tag{23}$$

$$\mathbf{H}_V \leftarrow \mathbf{H}_V \odot \frac{\mathbf{B}_V^T(\mathbf{Y} \oslash \hat{\mathbf{Y}})}{\mathbf{B}_V^T \mathbf{1}_{FT}} \tag{24}$$

where $\mathbf{1}_{MN}$ denotes an all-ones matrix composed of M rows and N columns, $\odot$ and $\oslash$ represent element-wise product and division, respectively.

These update rules are efficiently implemented and run iteratively until the cost function converges.

The implementation of the multiplicative update rules (20)-(24) has been performed following two parallelization techniques: (1) calling BLAS [44] and (2) using OpenMP directives. The multiplicative update rules results in matrix-matrix products (calculated by calling BLAS subroutine **dgemm**) along with other less computationally intensive auxiliary operations. Thus, the computational complexity of the parallel version is given by

$$O\left(\frac{FTKN_{\text{iter}}}{p}\right), \tag{25}$$

where $K$ represents the number of NMF bases and $N_{\text{iter}}$ is the total number of iterations of the NMF algorithm.

The pseudocode of this second stage can be founded in lines (5)-(13) of Algorithm 1.

### 3.3 Wheezing sound detection

Finally, wheezing detection is the last main stage of the proposed algorithm, as can be seen in Fig. 1. This stage determines the health status of the subject. To deal with this purpose, once obtained the spectral bases of the biomedical signal $\mathbf{B}_S$, the proposed algorithm has to differentiate between wheeze and normal breath sounds. This differentiation is carried out based on the periodic nature of wheeze sounds. In this sense, we propose to use a sparse descriptor, and particularly the Gini index $\beta$ in the frequency domain [37].

The Gini index $\beta$ computes the degree of periodicity of each basis estimated during the signal decomposition stage. A spectral basis with a high $\beta$ value denotes a high periodicity and can be classified as a wheezing basis. Mathematically, the Gini index descriptor is defined for the $k$-th spectral basis $\mathbf{b}_S(k) \in \mathbb{R}_+^F$ as follows

$$\beta(\mathbf{b}_S(k)) = \frac{F+1}{F} - \frac{2}{F} \frac{\sum_{f=1}^{F}(F+1-f)B_s(f,k)}{\sum_{f=1}^{F} B_S(f,k)} \tag{26}$$

where (26) is a normalized definition, i.e. $\beta(\mathbf{b}_S(k)) \in [0,1]$ for any spectral basis $\mathbf{b}_S(k)$.

After computing the Gini index for all bases $\mathbf{B}_S$, we propose to apply a thresholding process to cluster the bases. In particular, we fix a threshold $\gamma$ based on the median of the $\beta$ values provided by the Gini index for $\mathbf{B}_S$. Thus, we can identify the wheezing bases based on the criteria $\beta(\mathbf{b}_S(k)) \geq \gamma$ and arrange them in a new matrix $\mathbf{B}_W \in \mathbb{R}_+^{F \times K_W}$ where $K_W$ is the total number of wheezing bases (i.e., $K_W = K_S/2$). Similarly, $\mathbf{G}_W \in \mathbb{R}_+^{K_W \times T}$ is built based on the selected bases. Note that this thresholding approach has been widely exploited in the literature [45, 46] providing a promising discrimination performance between periodic and non-periodic bases.

To detect the patient's health status, the estimated wheezing spectrogram $\mathbf{X}_W \in \mathbb{R}_+^{F \times T}$ must be reconstructed, using the wheezing bases and their corresponding activations, as $\mathbf{X}_W = \mathbf{B}_W \mathbf{G}_W$. Then, the spectral energy distribution $\xi$ for $\mathbf{X}_W$ is computed as

$$\xi(f) = \sum_{t=1}^{T} X_W(f, t) \qquad (27)$$

In this manner, $\xi$ represents a vector composed of the spectral energy distribution along time frames. An unhealthy patient will present a spectral energy distribution concentrated in narrowband, i.e., a spectral peak from a periodic signal. Therefore, as in the previous case, we use the Gini index $\beta$ to determine whether the spectral energy distribution $\xi$ is related to an unhealthy or healthy patient [37] as follows

$$\Omega = \begin{cases} 1 & \text{if } \beta(\xi) \geq \gamma' \\ 0 & \text{if } \beta(\xi) < \gamma' \end{cases} \qquad (28)$$

where $\Omega \in \{0, 1\}$ labels the patient status as unhealthy ($\beta(\xi) \geq \gamma'$) or healthy ($\beta(\xi) < \gamma'$), and $\gamma'$ is a threshold fixed to 0.5 in order to guarantee the maximum detection rate [37].

Attending to the parallel implementation of the wheezing detection process, the computational complexity of this stage is given by

$$O\left(\frac{K_W F T}{p}\right). \qquad (29)$$

Note that this stage consists of a set of basic scalar operations and small search and sorting problems applied to independent vectors. Therefore, the solution adopted was to address these operations simultaneously by using the proper OpenMP directives (*pragma omp parallel for*) and the matrix operation by calling BLAS subroutine **dgemm**.

The pseudocode of the proposed wheezing detection algorithm is detailed in lines (14)-(21) of Algorithm 1.

---

**Algorithm 1** Pseudocode of the proposed multi-channel wheezing detection

    **Input**: Internal signal $x_I(n)$ and external signal $x_E(n)$.
1: Compute $\mathbf{X}_E$ and $\mathbf{X}_I$ using the STFT to obtain the signal representation in the frequency domain.
2: Compute the SVD of $\mathbf{X}_I$ to obtain the orthogonal matrices $\mathbf{U}$ and $\mathbf{V}$, respectively.
3: Initialize NMF basis matrices $\mathbf{B}_S$ and $\mathbf{B}_V$ using the matrix $\mathbf{U}$.
4: Initialize NMF gain matrices $\mathbf{G_S}$, $\mathbf{G_V}$ and $\mathbf{H}_V$ using random values.
5: **for** *iter* = 1 to $N_\text{iter}$ **do**
6:     Update $\mathbf{B}_S$ using (20).
7:     Update $\mathbf{B}_V$ using (21).
8:     Recompute the estimated signals in (6) and (5).
9:     Update $\mathbf{G}_S$ using (22).
10:     Update $\mathbf{G}_V$ using (23).
11:     Update $\mathbf{H}_V$ using (24).
12:     Recompute the estimated signals in (6) and (5).
13: **end for**
14: **for** $i$ = 1 to $k$ **do**
15:     Compute $\beta(\mathbf{b}_S(k))$ using (26).
16: **end for**
17: Compute $\gamma$ based on the median of the $\beta$ values provided by the Gini index for $\mathbf{B}_S$.
18: Cluster each basis of $\mathbf{B}_S$ using $\gamma$ threshold.
19: Reconstruct $\mathbf{X}_W$ using the clustering.
20: Compute $\xi(f)$ using (27).
21: Compute $\Omega$ using (28) to estimate the health status.
    **Output**: Health status.

---

## 4 Evaluation and experimental results

### 4.1 Experimental datasets

In this section, the proposed system is evaluated for the task of wheezing detection in multi-channel mixtures. In this evaluation, we have used two different types of experiments. First, we have assessed the performance and reliability of our proposal combining different datasets. As sound sources of ambient noises, we have used the dataset proposed in [14]. This audio collection takes into account a wide range of ambient noises classified as the most typical disturbing noises surrounding a medical office during the auscultation process. In this way, five types of ambient noise are considered: ambulance siren, baby crying, babble, car and street. As respiratory sound sources, we have combined the dataset proposed in [47] and the dataset proposed in [37]. The former is composed of 16 recordings of unhealthy and healthy patients, with duration between 4 and 51 s. The latter is composed of 8 mixtures from unhealthy patients with duration between 7 and 22 s, with a total of 41 wheezing and 63 respiratory cycles. To generate the multi-channel noisy mixtures, several Signal-to-Noise Ratios (SNR) [48] have been applied in the mixing process simulating high noisy environments.

The second experiment was conducted on a synthetic database to analyse the performance of the application in terms of efficiency and speedup. For this purpose, we generated multi-channel synthetic mixtures with different durations, from 60 to 900 s.

## 4.2 Experimental setup

The experimentation was conducted on the NVIDIA Jetson AGX Xavier development kit. This is an embedded system-on-chip (SoC) with an eight-core ARM v8.2 64-bit CPU. Xavier supports different kinds of running modes (configurable with the *NVPModel* command tool). In this way, different power consumption (10 W, 15 W, 30 W and full power), running cores (2, 4, 6 and 8) and CPU frequencies can be selected using *NVPModel*. This setup allows to emulate the upper bound for the best performance that can be reached with a wide range of mobile devices such as smartphones, laptops, tablets and other embedded systems under controlled conditions. Xavier runs Ubuntu Linux 18.04.1 LTS, the OpenBlas[1] library (release 0.3.20, February 2022), the FFTW[2] library (release 3.3.10, September 2021) and the GNU C Compiler 7 with the specification 4.5 of OpenMP. OpenBLAS is an optimized BLAS library based on GotoBLAS2 1.13 BSD.

To evaluate the detection performance, the sensitivity, the specificity and the accuracy metrics have been used. These metrics are commonly accepted in the field of wheezing detection and thus allow a fair comparison with other state-of-the-art methods. In particular, the sensitivity represents the probability of detecting wheezing samples correctly and can be computed as $SE = \frac{TP}{TP+FN}$. The specificity represents the probability of detecting normal breath samples correctly and can be computed as $SP = \frac{TN}{TN+FP}$. Finally, the accuracy represents the probability of detecting wheezing/normal breath samples correctly and can be computed as $ACC = \frac{TP+TN}{TP+FP+TN+FN}$. The terms $TP$, $FN$, $FP$ and $TN$ are the amount of the true positive, false negative, false positive and true negative test results, respectively.

To evaluate the benefits of proposal, the wheezing detection performance of our method has been compared with other state-of-the-art algorithms: HMMFL [47], TSVM [49] and MKNN [50]. MKNN and TSVM are supervised approaches. However, HMMFL and the proposed method do not use any type of training material.

| Methods | HMMFL | | | MKNN | | | TSVM | | | Proposal | | |
|---|---|---|---|---|---|---|---|---|---|---|---|---|
| SNR (dB) | SE (%) | SP (%) | ACC (%) | SE (%) | SP (%) | ACC (%) | SE (%) | SP (%) | ACC (%) | SE (%) | SP (%) | ACC (%) |
| -10 | **93.85** | 17.77 | 36.01 | 64.77 | 60.00 | 61.95 | 16.19 | **90.94** | 72.57 | 93.75 | 62.50 | **83.33** |
| -5 | **94.38** | 20.92 | 38.82 | 70.63 | 64.28 | 66.93 | 18.47 | **88.35** | 71.12 | 93.75 | 75.00 | **87.50** |
| 0 | 95.17 | 30.47 | 46.74 | 68.27 | 74.05 | 73.94 | 20.86 | **84.39** | 68.05 | **100.00** | 75.00 | **91.67** |
| 5 | 94.59 | 42.40 | 55.74 | 68.07 | 78.37 | 77.60 | 26.60 | 76.32 | 62.04 | **100.00** | 87.50 | **95.83** |
| 10 | 93.10 | 54.19 | 64.05 | 59.97 | 82.66 | 78.35 | 37.67 | 68.62 | 58.06 | **100.00** | 87.50 | **95.83** |

**Table 1**: Wheezing detection comparison between the proposed method and reference state-of-the-art methods.

## 4.3 Results

Firstly, Table 1 summarizes the comparison results between the proposed method and the reference state-of-the-art methods in terms of wheezing detection as a function of SNR. Note that Table 1 shows the mean results obtained for the different ambient noises. In general, the detection performance improves significantly as SNR increases. TSVM performs worse in terms of SE compared to SP. This means that this method has the ability to detect very clear wheezing sounds at the expense of others that are actually wheezing as well. Therefore, the number of FN is much higher than the number of FP. That could be due to the fact that this method is based on a cascade system composed of two SVM training models, where a sound is classified as wheezing if both models detect it. HMMFL has the opposite behavior to TSVM, i.e., the SP metric tends to be larger than the SE metric. This method has the ability to detect all wheezing sounds at the expense of wrongly annotating others. Therefore, the number of FP is much higher than the number of FN. Note that this method is based on Hidden Markov model (HMM) to detect frequency lines. This strategy fails when clinical noise is added in the same frequency band as the wheezing sounds, since the spectral lines detected are longer in duration than the wheezing sounds themselves. Finally, MKNN obtains intermediate results in terms of SE and SP.

On the other hand, the proposed method shows high robustness in terms of SE even in very adverse noisy environments (i.e., SNR = −10 dB). This demonstrates the strong ability of our proposal to detect the presence of wheezing sounds even when they may be masked by loud noises. Concerning the SP metric, the proposed method obtains worse results compared to SE. This suggests that the system tends to increase the number of FP in exchange for detecting the full range of wheezing sounds. In any case, for high SNR values, it obtains better results than the compared methods. Overall, the high ACC results reveals the great reliability of our proposal under adverse scenarios. According to the experiment, for SNR higher than 0 dB, the accuracy obtained by the proposal is always very similar, improving only a 4%. Note that for SNR = 5 *dB* and SNR = 10 *dB* only one audio sample was badly labeled. This is an important finding, because it reveals that the maximum detection performance is achieved in the SNR range [0, 5] dB.

---

[1] https://www.openblas.net
[2] http://www.fftw.org

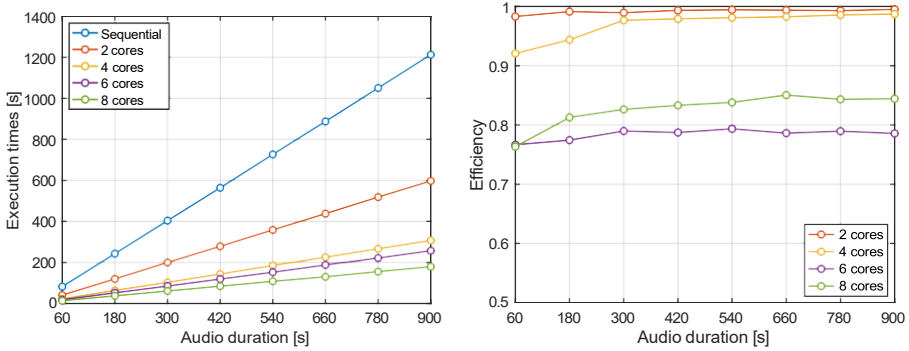

(a) Execution times measured in seconds.   (b) Evolution of the efficiency.

**Fig. 2**: Experimental results as a function of the audio length measured in seconds and the number of used cores of the NVIDIA AGX Xavier.

Secondly, a computational performance test has been designed and run. The computational results for this experiment are presented in Figure 2. Here, the length of the audio files was varied from 60 to 900 s in order to assess the efficiency and speedup of the proposal. Figure 2a shows the execution times as a function of the used cores of the Xavier device under full power conditions. As can be observed, the run time of the system increases with the length of the audio. In particular, the execution times of the sequential version are very high, which prevents its use in real applications. For example, for the case of audio samples with a duration of 15 minutes, the algorithm would run in more than 20 minutes. This fact justifies the use of parallel and high-performance techniques to solve the target problem.

Regarding the parallel approach, it can be seen that execution times decrease as the number of cores used increases. For the case of using 2 cores, the execution times are similar to the duration of the audio signals. However, in the other tested scenarios, the measured times are much shorter than the duration of the audios. Highlight that, in the case of 8 cores, the execution times are very low compared to the audio duration. For example, for the case of audio samples with a duration of 15 minutes, the algorithm would run in approximately three minutes.

The efficiency of the system is depicted in Figure 2b. As can be seen, efficiency has a very stable behavior even for short audio durations. Therefore, we can assert that our system scales correctly according to the theoretical complexity estimations when the number of processors and the size of the problem grow. In this regard, the OpenBLAS library has a weird behavior for 6-core runs. To verify that this behavior is only due to the combination of OpenBLAS and Xavier, several independent experiments were performed. These consisted of testing the performance of cblas_?gemm routines (where ? could be "d", "s", "c", "z", etc. depending on the data type used) with different types of matrices (square and rectangular). In this sense, for square matrices, the obtained efficiency for six cores was slightly superior to eight cores. For rectangular matrices, the performance for six cores was worse than for eight cores, especially when the dimensions of the resulting matrix were much smaller than the inner dimension of the matrix product. Note that the latter scenario is the most frequent in this proposal. Moreover, an independent test was also performed using OpenBLAS on a Xeon Silver 4110 processor. The behavior observed was very similar. Therefore, we can claim that this behavior is only due to how the library performs these operations.

| **Xavier mode** | Dur.[s] | SFFT | SVD | NMF | Detect. | **Total** |
|---|---|---|---|---|---|---|
| Mode 2 (4 cores, 15W) | 60  | 0.1 | 1.3  | 42.1  | 0.0 | 43.6  |
|                       | 180 | 0.2 | 2.6  | 123.3 | 0.1 | 126.3 |
|                       | 300 | 0.3 | 3.7  | 197.5 | 0.1 | 201.8 |
|                       | 420 | 0.5 | 5.6  | 276.2 | 0.2 | 282.5 |
|                       | 540 | 0.6 | 7.1  | 356.3 | 0.2 | 364.3 |
|                       | 660 | 0.7 | 8.1  | 439.3 | 0.3 | 448.5 |
|                       | 780 | 0.8 | 9.9  | 521.1 | 0.3 | 532.2 |
|                       | 900 | 1.0 | 10.8 | 601.1 | 0.4 | 613.4 |
| Mode 3 (8 cores, 30W) | 60  | 0.1 | 1.1 | 24.0  | 0.0 | 25.3  |
|                       | 180 | 0.2 | 1.9 | 69.1  | 0.1 | 71.3  |
|                       | 300 | 0.3 | 2.7 | 115.0 | 0.1 | 118.1 |
|                       | 420 | 0.3 | 3.6 | 161.3 | 0.1 | 165.4 |
|                       | 540 | 0.4 | 4.3 | 212.4 | 0.2 | 217.3 |
|                       | 660 | 0.5 | 4.8 | 263.6 | 0.2 | 269.1 |
|                       | 780 | 0.5 | 5.6 | 312.4 | 0.2 | 318.9 |
|                       | 900 | 0.6 | 6.7 | 357.0 | 0.3 | 364.7 |
| 4 cores, full power   | 60  | 0.1 | 0.5 | 21.7  | 0.0 | 22.4  |
|                       | 180 | 0.1 | 1.3 | 62.8  | 0.0 | 64.3  |
|                       | 300 | 0.2 | 1.9 | 101.1 | 0.1 | 103.3 |
|                       | 420 | 0.3 | 2.5 | 141.2 | 0.1 | 144.1 |
|                       | 540 | 0.3 | 2.9 | 181.8 | 0.1 | 185.1 |
|                       | 660 | 0.4 | 3.7 | 221.4 | 0.1 | 225.7 |
|                       | 780 | 0.5 | 4.3 | 261.5 | 0.2 | 266.5 |
|                       | 900 | 0.5 | 5.0 | 301.3 | 0.2 | 307.0 |
| 8 cores, full power   | 60  | 0.1 | 0.6 | 12.8  | 0.0 | 13.5  |
|                       | 180 | 0.2 | 0.9 | 36.2  | 0.0 | 37.3  |
|                       | 300 | 0.2 | 1.4 | 59.3  | 0.1 | 61.1  |
|                       | 420 | 0.3 | 1.7 | 82.6  | 0.1 | 84.7  |
|                       | 540 | 0.3 | 2.0 | 105.9 | 0.1 | 108.4 |
|                       | 660 | 0.3 | 2.6 | 127.4 | 0.1 | 130.4 |
|                       | 780 | 0.3 | 2.9 | 152.4 | 0.1 | 155.7 |
|                       | 900 | 0.3 | 3.3 | 175.7 | 0.2 | 179.5 |

**Table 2**: Execution times measured in seconds for each stage and operating mode.

Finally, as the purpose is to test our system on up-to-date embedded systems, we have selected two test modes of Xavier following the current market trend: Mode 2 (4 cores, 15 W) and Mode 3 (8 cores, 30 W). Table 2 shows the computational results obtained for the algorithm stages and for these operation modes compared to the full power modes. Again, it can be observed that the execution times of Mode 2 are a little below the length of the audios, while in the case of Mode 3 they remain significantly below this length (being almost a third). Focusing on the stages that compound the system, it can be seen that the running times of the signal representation and wheezing detection stages are negligible in relation to the total execution time. On the other hand, the measured times from the SVD stage are approximately two order of magnitude lower than those of the signal decomposition stage.

| Xavier mode | Dur.[s] | SFFT | SVD | NMF | Detect. | Total |
|---|---|---|---|---|---|---|
| Mode 2 (4 cores, 15W) | 60 | 0.55 | 0.19 | 0.94 | 0.51 | 0.93 |
|  | 180 | 0.76 | 0.18 | 0.95 | 0.44 | 0.95 |
|  | 300 | 0.82 | 0.20 | 0.99 | 0.51 | 0.98 |
|  | 420 | 0.77 | 0.18 | 0.99 | 0.54 | 0.98 |
|  | 540 | 0.82 | 0.18 | 0.98 | 0.54 | 0.98 |
|  | 660 | 0.84 | 0.19 | 0.97 | 0.55 | 0.98 |
|  | 780 | 0.83 | 0.18 | 0.97 | 0.54 | 0.98 |
|  | 900 | 0.85 | 0.19 | 0.97 | 0.48 | 0.98 |
| Mode 3 (8 cores, 30W) | 60 | 0.29 | 0.12 | 0.84 | 0.34 | 0.81 |
|  | 180 | 0.50 | 0.14 | 0.86 | 0.37 | 0.84 |
|  | 300 | 0.50 | 0.14 | 0.85 | 0.35 | 0.83 |
|  | 420 | 0.58 | 0.14 | 0.85 | 0.37 | 0.83 |
|  | 540 | 0.65 | 0.13 | 0.83 | 0.31 | 0.82 |
|  | 660 | 0.66 | 0.15 | 0.81 | 0.34 | 0.80 |
|  | 780 | 0.64 | 0.15 | 0.82 | 0.35 | 0.80 |
|  | 900 | 0.70 | 0.15 | 0.82 | 0.34 | 0.81 |
| 4 cores, full power | 60 | 0.49 | 0.27 | 0.94 | 0.47 | 0.92 |
|  | 180 | 0.61 | 0.22 | 0.96 | 0.55 | 0.94 |
|  | 300 | 0.73 | 0.23 | 0.99 | 0.46 | 0.98 |
|  | 420 | 0.73 | 0.24 | 0.99 | 0.55 | 0.98 |
|  | 540 | 0.77 | 0.25 | 0.99 | 0.53 | 0.98 |
|  | 660 | 0.78 | 0.25 | 0.99 | 0.53 | 0.98 |
|  | 780 | 0.82 | 0.26 | 0.99 | 0.53 | 0.99 |
|  | 900 | 0.83 | 0.23 | 0.99 | 0.54 | 0.99 |
| 8 cores, full power | 60 | 0.31 | 0.13 | 0.80 | 0.27 | 0.76 |
|  | 180 | 0.28 | 0.16 | 0.83 | 0.32 | 0.81 |
|  | 300 | 0.35 | 0.15 | 0.85 | 0.29 | 0.83 |
|  | 420 | 0.37 | 0.17 | 0.85 | 0.33 | 0.83 |
|  | 540 | 0.50 | 0.18 | 0.85 | 0.33 | 0.84 |
|  | 660 | 0.62 | 0.18 | 0.86 | 0.36 | 0.85 |
|  | 780 | 0.64 | 0.19 | 0.86 | 0.35 | 0.84 |
|  | 900 | 0.63 | 0.18 | 0.86 | 0.33 | 0.84 |

**Table 3**: Efficiency for each stage and operating mode.

An interesting conclusion of the effect of reducing the energy consumption of the platform could be inferred from the results obtained. As can be observed, execution times are reduced by half when no power limits are imposed.

Table 3 shows the efficiency evolution obtained for the algorithm stages. Note that the efficiency for Mode 2 was computed with respect to the sequential version limited to 15 W, the efficiency for Mode 3 was computed with respect to the sequential version limited to 30 W, and the full power modes efficiencies were computed with respect to the sequential version without power limit. As expected, limiting the power of Xavier has no impact on the efficiency of the algorithm. Finally, as can be seen in the results obtained, the overall efficiency is determined by the efficiency of the NMF stage, which is the one with the highest computational burden.

# 5 Conclusion

Auscultation has proven to be a very useful procedure to diagnose cardio-respiratory pathologies. However, this technique suffers from two main limitations, the high dependence on ambient noise surrounding auscultation and the high subjectivity of physicians. Most recent approaches have shown ways to overcome some of the drawbacks of this medical technique, although with evident limitations. The work proposed in this paper tackles the compound problem of the background denoising and the wheezing detection from a multi-channel recording captured during the auscultation process, providing a real-time implementation. To our best knowledge, our proposal is the first implementation that addresses this problem obtaining reliable results along with promising computation times. This has been achieved by the intensive use of parallel and high-performance computation techniques. In particular, the proposal is decomposed into three main stages: signal preprocessing, signal decomposition and wheezing detection. The signal preprocessing stage deals with the representation of the input signal and with the parameter initialization. Parameter initialization is based on SVD and tries to avoid dependence on the initial values of NMF parameters. The decomposition stage implements the NMF rules for the decomposition of the multi-channel approach. Finally, the wheezing detection is performed based on the Gini index $\beta$ sparse descriptor in the frequency domain.

Finally, for future work, the challenge would be to design a hardware prototype to implement the proposed method. In addition, biomedical features other than wheezing would be analyzed to develop a framework capable of detecting multiple pathologies.

# Acknowledgments

This work was supported by MCIN/AEI/10.13039/501100011033 under the projects grant *PID2020-119082RB-{C21,C22}*, by Gobierno del Principado de Asturias under grant *AYUD/2021/50994*, by the Programa Operativo FEDER Andalucia 2014-2020 under project with reference *1257914* and by the Ministry of Economy, Knowledge and University, Junta de Andalucia under Project *P18-RT-1994*.

# Data availability

Data generated during the current study are available from the corresponding author on reasonable request.

# Conflict of interest

The authors declare that there is no conflict of interest.

# References

[1] Abbas, A. K. & Bassam, R. Phonocardiography signal processing. *Synthesis Lectures on Biomedical Engineering* **4** (1), 1–194 (2009) .

[2] Sarkar, M., Madabhavi, I., Niranjan, N. & Dogra, M. Auscultation of the respiratory system. *Annals of thoracic medicine* **10** (3), 158 (2015) .

[3] Kumar, D., Carvalho, P., Antunes, M. & Henriques, J. Noise detection during heart sound recording 3119–3123 (2009). https://doi.org/10.1109/IEMBS.2009.5332569 .

[4] Taplidou, S. A. & Hadjileontiadis, L. J. Wheeze detection based on time-frequency analysis of breath sounds. *Computers in biology and medicine* **37** (8), 1073–1083 (2007) .

[5] Suzuki, A., Sumi, C., Nakayama, K. & Mori, M. Real-time adaptive cancelling of ambient noise in lung sound measurement. *Medical and Biological Engineering and Computing* **33** (5), 704–708 (1995) .

[6] Patel, S. B. *et al.* An adaptive noise reduction stethoscope for auscultation in high noise environments. *The Journal of the Acoustical Society of America* **103** (5), 2483–2491 (1998) .

[7] Fleeter, J. S. & Wodicka, G. R. Auscultation of heart and lung sounds in high-noise environments using adaptive filters. *The Journal of the Acoustical Society of America* **104** (3), 1781–1781 (1998) .

[8] Della Giustina, D., Riva, M., Belloni, F. & Malcangi, M. Embedding a multichannel environmental noise cancellation algorithm into an electronic stethoscope (2011) .

[9] Nelson, G., Rajamani, R. & Erdman, A. Noise control challenges for auscultation on medical evacuation helicopters. *Applied Acoustics* **80**, 68–78 (2014) .

[10] Evans, N. W., Mason, J. S., Liu, W. M. & Fauve, B. *An assessment on the fundamental limitations of spectral subtraction* Vol. 1 (2006).

[11] Chang, C.-C. *et al.* Regulation of metastatic ability and drug resistance in pulmonary adenocarcinoma by matrix rigidity via activating c-met and egfr. *Biomaterials* **60**, 141–150 (2015) .

[12] Emmanouilidou, D., McCollum, E. D., Park, D. E. & Elhilali, M. Adaptive noise suppression of pediatric lung auscultations with real applications to noisy clinical settings in developing countries. *IEEE Transactions on Biomedical Engineering* **62** (9), 2279–2288 (2015) .

[13] Emmanouilidou, D., McCollum, E. D., Park, D. E. & Elhilali, M. Computerized lung sound screening for pediatric auscultation in noisy field environments. *IEEE Transactions on Biomedical Engineering* **65** (7), 1564–1574 (2017) .

[14] De La Torre Cruz, J. *et al.* An incremental algorithm based on multichannel non-negative matrix partial co-factorization for ambient denoising in auscultation. *Applied Acoustics* **182**, 108229 (2021). URL https://www.sciencedirect.com/science/article/pii/S0003682X21003236. https://doi.org/https://doi.org/10.1016/j.apacoust.2021.108229 .

[15] Xie, Y., Xie, K., Yang, Q. & Xie, S. Reverberant blind separation of heart and lung sounds using nonnegative matrix factorization and auxiliary function technique. *Biomedical Signal Processing and Control* **69**, 102899 (2021) .

[16] Muñoz-Montoro, A. J., Revuelta-Sanz, P., Villalón-Fernández, A., Muñiz, R. & Ranilla, J. A system for biomedical audio signals processing based on high performance computing techniques (2022) .

[17] Fraiwan, L. *et al.* Automatic identification of respiratory diseases from stethoscopic lung sound signals using ensemble classifiers. *Biocybernetics and Biomedical Engineering* **41** (1), 1–14 (2021). URL https://www.sciencedirect.com/science/article/pii/S0208521620301297. https://doi.org/https://doi.org/10.1016/j.bbe.2020.11.003 .

[18] Pasterkamp *et al.* Respiratory sounds: ad vances beyond the stethoscope. *American journal of respiratory and critical care medicine* **156 (3)**, 974–987 (1997) .

[19] Sovijarvi *et al.* Definition of terms for applications of respiratory sounds. *Eu ropean Respiratory Review* **10 (77)**, 597–610 (2000) .

[20] Lin *et al.* Wheeze recognition based on 2d bilateral filtering of spectrogram. *Biomedical Engineering: Applica tions Basis and Communications* **18 (03)**, 128–137 (2006) .

[21] Wisniewski, M., Zielinski & P., T. Fast and robust method for wheezes recog nition in remote asthma monitoring. *In Information technologies in biomedicine. Springer.* 568–576 (2012) .

[22] Torre-Cruz *et al.* Wheezing sound separation based on constrained nonnegative matrix factorization. *Proceedings of the 10th international conference on bioinformatics and biomedical technology, ACM* 18−-24 (2018) .

[23] Rocha, B. M., Pessoa, D., Marques, A., Carvalho, P. & Paiva, R. P. Influence of event duration on automatic wheeze classification 7462–7469 (2021). https://doi.org/10.1109/ICPR48806.2021.9412226 .

[24] Alic *et al.* A novel approach to wheeze detection. *In World congress on medical physics and biomedical engineering* 963–966 (2007) .

[25] Mendes *et al.* Detection of wheezes using their signature in the spectrogram space and musical features **In 37th annual international conference of the IEEE engineering in medicine and biology society**, 5581–5584 (2015) .

[26] Oletic *et al.* Lowpower wearable respiratory sound sensing. *Sensors* **14 (4)**, 6535–6566 (2014) .

[27] Mayorga *et al.* Acous tics based assessment of respiratory diseases using gmm classification. *Annual international conference of the IEEE engineering in medicine and biology* 6312–6316 (2010) .

[28] Qiu *et al.* Automatic wheeze detection based on auditory modelling. *Proceedings of the Institution of Mechanical Engineers Part H: Journal of Engineering in Medicine* **219 (3)**, 219–227 (2005) .

[29] Le Cam, S., Belghith, A., Collet, C. & Salzenstein, F. Wheezing sounds detection using multivariate generalized gaussian distributions. *In IEEE international conference on acoustics, speech and signal processing* 541–544 (2009) .

[30] Cortes *et al.* Monitoring of wheeze duration during spontaneous respiration in asthmatic patients. *In 27th annual interna tional conference of the ieee engineering in medicine and biology society* 6141–6144 (2006) .

[31] Lin *et al.* Automatic wheezing detection based on signal processing of spectrogram and backpropagation neural network. *Journal of healthcare engineering* **6 (4)**, 649−-672 (2015) .

[32] Kochetov *et al.* Wheeze detection using convolutional neural networks. *In EPIA conference on artificial intelligence, Springer* 162–173 (2017) .

[33] Kuo, H.-C., Lin, B.-S., Wang, Y.-D. & Lin, B.-S. Development of automatic wheeze detection algorithm for children with asthma. *IEEE Access* **9**, 126882–126890 (2021) .

[34] Semmad, A. & Bahoura, M. Serial hardware architecture of multilayer neural network for automatic wheezing detection 28–31 (2021). https://doi.org/10.1109/MWSCAS47672.2021.9531850 .

[35] Semmad, A. & Bahoura, M. Long short term memory based recurrent neural network for wheezing detection in pulmonary sounds 412–415 (2021). https://doi.org/10.1109/MWSCAS47672.2021.9531784 .

[36] Torre-Cruz *et al.* A novel wheezing detection approach based on constrained non-negative matrix factorization. *Applied Acoustics* **148**, 276–288 (2019) .

[37] Torre-Cruz, J., Cañadas Quesada, F. J., Carabias Orti, J. J., Vera Candeas, P. & Ruiz Reyes, N. Combining a recursive approach via nonnegative matrix factorization and gini index sparsity to improve reliable detection of wheezing sounds. *Expert Systems with Applications* **147**, 113212 (2020). URL https://www.sciencedirect.com/science/article/pii/S0957417420300385. https://doi.org/10.1016/j.eswa.2020.113212 .

[38] Boutsidis, C. & Gallopoulos, E. Svd based initialization: A head start for nonnegative matrix factorization. *Pattern Recognition* **41** (4), 1350–1362 (2008). URL https://www.sciencedirect.com/science/article/pii/S0031320307004359. https://doi.org/https://doi.org/10.1016/j.patcog.2007.09.010 .

[39] Stewart, G. W. Perturbation theory for the singular value decomposition. Tech. Rep. (1998).

[40] Frigo, M. & Johnson, S. G. The design and implementation of fftw3. *Proceedings of the IEEE* **93** (2), 216–231 (2005) .

[41] Dagum, L. & Menon, R. Openmp: an industry standard api for shared-memory programming. *IEEE computational science and engineering* **5** (1), 46–55 (1998) .

[42] Dongarra, J. *et al.* The singular value decomposition: Anatomy of optimizing an algorithm for extreme scale. *SIAM review* **60** (4), 808–865 (2018) .

[43] Févotte, C., Bertin, N. & Durrieu, J.-L. Nonnegative matrix factorization with the itakura-saito divergence: With application to music analysis. *Neural computation* **21** (3), 793–830 (2009) .

[44] Blackford, L. S. *et al.* An updated set of basic linear algebra subprograms (blas). *ACM Transactions on Mathematical Software* **28**, 135–151 (2001) .

[45] Toh, K. K. V. & Isa, N. A. M. Noise adaptive fuzzy switching median filter for salt-and-pepper noise reduction. *IEEE signal processing letters* **17** (3), 281–284 (2009) .

[46] Rafii, Z. & Pardo, B. Repeating pattern extraction technique (repet): A simple method for music/voice separation. *IEEE transactions on audio, speech, and language processing* **21** (1), 73–84 (2012) .

[47] Oletic, D. & Bilas, V. Asthmatic wheeze detection from compressively sensed respiratory sound spectra. *IEEE journal of biomedical and health informatics* **22** (5), 1406–1414 (2017) .

[48] Tucker, D. G. & Gazey, B. K. *Applied underwater acoustics* (Elsevier Science & Technology, 1966).

[49] Mazi´c, I., Bonkovi´c, M. & Džaja, B. Two-level coarse-to-fine classification algorithm for asthma wheezing recognition in children's respiratory sounds. *Biomedical Signal Processing and Control* **21**, 105–118 (2015) .

[50] Shaharum, S. M., Sundaraj, K., Aniza, S., Palaniappan, R. & Helmy, K. Classification of asthma severity levels by wheeze sound analysis 172–176 (2016). https://doi.org/10.1109/SPC.2016.7920724 .